\documentstyle[eqsecnum,preprint,tighten,aps]{revtex}
\newfont{\largemi}{cmmi10}
\newfont{\smallmi}{cmmi6}

\begin{document}
\draft

\title{A simple approach to the angular momentum distribution in the
ground states  of many-body systems }

\author{ Y. M. Zhao$^{a,c}$
 A. Arima$^{b}$, and N. Yoshinaga$^{a}$}

\vspace{0.2in}
 \address{$^a$ Department of Physics,
 Saitama University, Saitama-shi, Saitama 338 Japan \\
$^b$ The House of Councilors, 2-1-1 Nagatacho, 
Chiyodaku, Tokyo 100-8962, Japan \\
$^c$ Department of Physics,  Southeast University, Nanjing 210018 China }

\date{\today}
\maketitle

\begin{abstract}

We propose  a simple  approach to predict the angular momentum $I$
ground state ($I$g.s.) probabilities of many-body systems that
does not require the diagonalization of hamiltonians with random
interactions. This method is  found to be applicable to {\bf all}
cases that have been discussed: even and odd fermion systems (both
in single-$j$ and many-$j$ shells), and boson (both $sd$ and
$sdg$) systems. A simple relation for the highest angular momentum
 g.s. probability
 is  found.  Furthermore, it is suggested  for the first time
 that the 0g.s. dominance in boson systems
 and in even-fermion systems is   given by  two-body
 interactions with specific features.

\end{abstract}

\pacs{PACS number:   05.30.Fk, 05.45.-a, 21.60Cs, 24.60.Lz}

\vspace{0.4in}
        
\newpage

The low-lying spectra of many-body systems with even
 numbers of fermions  were recently examined by
 Johnson {\it et al.} \cite{Johnson1,Johnson2,Johnson3}
using  the two-body random ensemble (TBRE), and the results showed
a preponderance of $I^{\pi}=0^+$ ground states (0g.s.). The 0g.s.
dominance  was soon confirmed in  $sd$-boson systems
\cite{Bijker,Casten}. Therefore, the 0g.s. dominance in even
fermion systems and boson systems is  robust and insensitive to
the detailed  statistical properties of the random hamiltonian,
suggesting that the  {\em pairing} features arise from a very
large ensemble of two-body interactions
other than a simple monopole pairing interaction  
 and might be independent
of the specific character of the force.
 Very recently, several interesting
studies were performed to check whether the spectroscopy based
random and/or displaced random ensembles can simulate that of
realistic systems \cite{Horoi,Zhao0,Zuker}.

An understanding of the 0g.s. dominance in many-body systems is
extremely important, since this observation seems to be contrary
to what is traditionally assumed. For example, in nuclear physics
the 0g.s. dominance in even-even nuclei is usually explained as a
reflection of the strong pairing that results for a short-range
attraction between identical nucleons.

There have been several efforts to understand this observation.
Work has been reported on its connection with the distribution
widths of eigenvalues \cite{Bijker1} or the distribution of the
lowest eigenvalues of each angular momentum \cite{Bijker}, its
connection with geometric chaoticity of the angular momentum
coupling \cite{zelevinsky}, its connection with the largest and
smallest diagonal matrix elements \cite{Zhaox}, its connection
with random polynomials \cite{Kus} or a Hartree-Bose mean field
analysis\cite{Bijkerx}, and its connection with deviations from
the random theory expectations \cite{Johnsonx}. These studies are
interesting and important, and have potentially impacted our
understanding on the origin of one of the most characteristic
features of nuclear spectra. All of these approaches, however,
address only simple or very specific ($sp$ and $sd$ bosons, or
fermions in small and single-$j$ shells) cases. It is desirable,
therefore,  to construct a {\it simple and  universal} approach to
understand the 0g.s. dominance of even-fermion systems and
$sd$-boson systems,  which can also be applied when discussing the
probability that the g.s. has angular momentum $I$ ($I$g.s. probability)
 of very different systems (boson systems, even and 
 odd fermions in single-$j$ or many-$j$ shells.).

Towards that goal, we present in this Letter a new view on the
origin of the 0g.s. dominance in even-fermion and boson systems.
For the first time, the $I$g.s. probabilities of both fermions (in
single-$j$ and many-$j$ shells) and bosons
 are  studied on the same footing.   In
the process, some previously unrecognized features of the highest 
angular momentum $I=I_{max}$g.s. probabilities of
  fermions in single-$j$ and many-$j$ shells,  and of $sd$ and $sdg$ bosons
will be discerned and explained, thereby providing further new
insights on the problem. We shall show that the 0g.s. dominance in
even-fermion and boson systems   comes from
two-body interactions with specific features.

We now very succinctly describe our method and then discuss its
application to the various possible cases. The basic idea is as
follows. We first set one of the two-body matrix elements of the
problem to -1 and all the rest to zero and then see which  angular
momentum $I$ gives the lowest eigenvalue among {\bf all} of the
eigenvalues of this many-body system. If the number of independent
two-body interaction matrix elements is $N$, the above procedure
is iterated $N$ times, with each of the matrix elements assuming
the privileged role of being set to -1. After all $N$ calculations
have been done, we simply count how many times (denoted as ${\cal
N}_I$) the angular momentum $I$ gives the lowest eigenvalue
\cite{note1}. Finally, the $I$g.s. probability  
  is given by ${\cal N}_I/N$.

We will now show the universality of this simple method by
comparing its predictions with the results obtained  by
diagonalizing random hamiltonians for a variety of very different
systems. In all cases, we will use the
TBRE\cite{Johnson1,Johnson2,Bijker,Casten,Zhaox,Kus} to define
the random two-body interactions, $G_J(j_1j_2j_3j_4)$. Namely,
$G_J(j_1j_2j_3j_4)$ will be chosen to be random numbers with a
distribution
\begin{equation}
  \rho(G_J(j_1j_2j_3j_4)) =
  \frac{1}{\sqrt{2\pi}} {\rm exp} \left(-\frac{\left(G_J(j_1j_2j_3j_4)\right)^2}{2} \right).   \label{tbre}
\end{equation}
For single-$j$ shells, the labels $(j_1j_2j_3j_4)$ are unnecessary
and will thus be suppressed.

Let us start by considering fermions in single-$j$ shells. TABLE I
shows the angular momenta $I$ which produce the lowest
eigenvalues for the different possible choices for which $G_J$ to be 
set to -1. The results are shown for $j$ values ranging from
$\frac{7}{2}$ to $\frac{31}{2}$ and for systems of $n=4$.
The number ${\cal N}_{I=0}$ of 0g.s staggers with $j$
with an interval $\delta j$=3. FIG. 1 gives a comparison between
the predicted 0g.s. probabilities (open diamond symbols) and those
obtained by diagonalizing the TBRE hamiltonian (solid square
symbols). The agreement is quite good, both for small-$j$ and
large-$j$ shells. The predicted 0g.s. probabilities show the same
staggering character as  those obtained using the TBRE hamiltonian. A similar
staggering behavior of 0g.s. probabilities for  single-$j$ shells
with $n= 6$ was observed earlier \cite{Zhao0} and it can be
explained in the same way .

In  single-$j$ shells, the highest angular momentum state (denoted
as $I_{max}$) was found to have a sizable probability to be the
g.s. \cite{zelevinsky,Zhaox}. This  can be understood from the
observation that ${\cal N}_{I_{max}} =1$ always. It is easy to
confirm that the eigenvalue of the $I=I_{max}$ state is the
lowest when $G_{J_{max}} = -1$ and all other parameters are
switched off. A detailed study of the  eigenvalue of the
state with $I=I_{max}$ in single-$j$ shells was given in Ref.
\cite{Zhao0}. Because  ${\cal N}_{I_{max}} =1$, the predicted
$I_{max}$g.s. probabilities
 of fermions in single-$j$ shells are
$\frac{1}{N} = \frac{1}{j+1/2} \times 100 \%$, a formula which is
valid for all particle numbers (even or odd). It is predicted that
the $I=I_{max}$g.s. probability  decreases gradually with $j$ and
vanishes in the large-$j$ limit.

In Ref. \cite{Bijker} the authors found that the $I_{max}$g.s.
probability is also large for $sd$-boson systems. This can be
explained in the same way. Among the  two-body interactions, the
interaction $- (d^{\dagger} d^{\dagger})^4 (d d)^4$ always gives
the lowest eigenvalue for the $I_{max}=2n$ state, independent of
the boson number.In the $sd$-boson model, the predicted $I=2n$
g.s. probability is 1/7=14.3$\%$, well consistent \cite{note2}
 with the previous results
($\sim 15\%$)  \cite{Bijker,Zhao0}.

FIG. 2a) shows the  $I_{max}$ probabilities for different fermion 
numbers in   single-$j$ shells.  The $I_{max}$g.s.  probabilities
obtained by diagonalizing the TBRE hamiltonian and those based
on our simple $\frac{1}{N} \times 100\%$ ($N=(2j+1)/2$) rule are in  good
agreement. For the first time, the $I_{max}$g.s. probabilities of
fermions in single-$j$ shells are shown to be independent of
particle number $n$  and to follow a simple analytic relation.

Our argument on the $I_{max}$g.s. probabilities for single-$j$
shells can be readily generalized to many-$j$ shells. Consider,
for example, two shells with angular momenta $j_1$ and $j_2$.
Following the same logic as was used for a single-$j$ shell, we
predict that the two angular momenta $I_{max}'=I_{max}(j_1^n)$ and
$I_{max}(j_2^n)$ have  g.s. probabilities which are at least as
large as $1/N \times 100\%$. Here, $I_{max}(j^n)$ refers to the
highest angular momentum of a state constructed from the $j^n$
configuration. In other words, we can predict in this way the
lower limit on these $I_{max}'$g.s. probabilities. This is
analogous to the Schmidt diagram of magnetic moments of
odd-$A$ nuclei in nuclear structure theory.

FIG. 2b) presents the $I_{max}'=I_{max}(j_1^n)$ and $I_{max}(j_2^n)$
g.s. probabilities. They are compared with a simple $1/N$ plot.
Here $N$ is the number of independent  two-body interactions
of a ($j_1, j_2$) shell. 
Indeed, the $1/N$ predicted lower limit on the $I_{max}'$g.s.
probabilities works quite well. It should also be noted that the
$I$g.s. probabilities with $I$ very near $I_{max}'$ are extremely
small (less than 1$\%$) in these examples.

FIGS. 3a) and 3b) present a comparison between  the predicted $I$g.s.
probabilities and those obtained by diagonalizing the TBRE
hamiltonian for a $j=\frac{9}{2}$ shell. We present two cases: the
case of  $n=4$ (even) is shown in 3a) and the case of $n=5$ (odd) is
shown in 3b). In both cases, the agreement is good. Note that the
agreement does not deteriorate  when we go to
larger single-$j$ shells or to many-$j$ shells where there are
more interaction parameters.

FIGS. 3c) and 3d) show
the $I$g.s. probabilities for
a system of 7 fermions in two-$j$
shells ($j_1=\frac{7}{2},~j_2 = \frac{5}{2}$), and 
an $sd$-boson
system (with 10 bosons),
respectively. The
predicted $I$g.s. probabilities are reasonably consistent with
those obtained by diagonalizing the TBRE hamiltonian. We checked
many cases, such as (a) single-$j$ shells with $j$ up to
$\frac{31}{2}$ and with $n$=4, 5, 6, (b) two-$j$ shells with
$(2j_1,~2j_2)$=$(5,~7)$, $(5,~9)$, ($11,~3)$, $(11,~5)$, $(11,~9)$
and $(13,~9)$ and with $n=4, ~5, ~6$, (c) $sd$-boson systems with
$n$ up to 17,  and (d) $sdg$-boson systems with $n=4,~5$, and $6$,
and in {\bf all} cases the agreement is reasonably  good.

One may ask why we study the  $I$g.s. probabilities
in the way proposed in this Letter.
A  rationale can be seen from the following analysis.  
The eigenvalues, though non-linear in
principle, are linear in terms of each interaction in a local space
in which we set the one non-zero matrix element to
-1 and looked for the state with the lowest eigenvalue.  
Therefore, instead of studying the effects of all two-body matrix
elements simultaneously, we decompose the problem into $N$ parts.
In each part, we focus on only one interaction matrix element.
 As a specific example, we
shall consider a single-$j$ shell with $j=\frac{31}{2}$. We set
$G_0=-1$ and all other matrix elements to their TBRE values {\em
multiplied by a factor $\epsilon$}, with $\epsilon$ running from
$0$ to $10$. As expected, almost all of the ground states have
$I=0$ when $\epsilon$ is small.  Interestingly, the 0g.s.
probability remains large even when $\epsilon$ becomes large (say,
1.5). If we switch off all of the interaction matrix elements for
which the lowest eigenvalue corresponds to an $I=0$ state, then
the 0g.s. probabilities become small. Similar results can be
noticed in other cases.

 Previously, it was noticed that
 the 0g.s. dominance is not dependent on having monopole pairing
 for fermions in the $sd$-shell \cite{Johnson1,Johnson2,Johnson3}.
 It was not known, however, which interactions
 are crucial in order to have 0g.s. dominance, and it was assumed
 by many authors that 0g.s. dominance is thus an intrinsic
 property of the model space.
Using the method proposed in this Letter, we are able to readily
tell which interactions (not only monopole pairing interactions)
 are crucial for 0g.s. dominance.

To summarize, we have presented in this Letter a simple approach
to predict the $I$g.s. probabilities for  many-body systems. The
agreement between the predicted $I$g.s. probabilities and those
obtained by diagonalizing the TBRE random hamiltonian  
 is good. This method is applicable to both (even and odd)
fermion systems (with both single-$j$ and multi-$j$ shells)  and
to boson systems. It predicts the 0g.s. probability  and addresses
the other $I$g.s. probabilities as well. We believe, therefore,
that we have provided in this Letter a universal approach for
studying the $I$g.s. probabilities.

Using this method, we have been able to address several important
issues regarding the spectra of random hamiltonians.  
We have shown, for the first time,  that the  
$I_{max}$g.s. probabilities of fermions in single-$j$ shells are
determined solely by the number of two-body matrix elements  and
are independent of particle number, and furthermore that they
follow a simple $1/N$ relation. A generalization of such a
regularity to fermion systems in many-$j$ shells  and to ($sd$ and
$sdg$) boson systems has been shown to work well too. In addition,
we discovered which interactions (not only the monopole
pairing interaction) are essential for
producing 0g.s. dominance in  boson and
even-fermion systems. Our
results suggest that the 0g.s. dominance (and other $I$g.s.
dominance in odd-A valence fermion systems) in boson systems and
even-fermion systems is not a reflection of an intrinsic property 
of the model space, as
previously assumed, but is related to the two-body matrix elements
that give $I=0$ ground states when they are set to -1 and all
others are set to 0. The analogous remark applies to the dominance
of a given $I$g.s in odd-A fermion systems.

What is not yet understood at a more microscopic level is why
${\cal N}_0$ is so large for even-fermion systems, nor why there
is a staggering of the ${\cal N}_0$ values for an even number of
fermions in single-$j$ shells. Further consideration of these
issue is warranted.

We are grateful to Drs. S. Pittel, R. F. Casten, and
R. Bijker  for discussions and communications.
This work is  supported in part by the Japan Society
for the Promotion of Science
(contract ID: P01021).

\newpage

{\bf Caption}:

FIG. 1 ~~ The $I=$0
  g.s. probabilities of 4 fermions in 
different single-$j$ shells. Solid
squares
 are obtained from 1000 runs of the TBRE, 
and the open diamonds are from our simple prescription.

\vspace{0.2in}

FIG. 2 ~(a) Calculated $I_{max}$g.s. probabilities for single-j
shells. The solid line plots the results from our simple
prescription whereas the other results were obtained by
diagonalizing  the TBRE hamiltonian. (b). The
$I_{max}'=I_{max}(j_1^n)$ and $I_{max}(j_2^n)$
 g.s. probabilities for two-$j$
shells.  The lower limit of the
 $I_{max}'$ g.s. probabilities is predicted by  our simple formula
  $\frac{1}{N} \times 100\%$
(solid line). The other results were  obtained by 
diagonalizing  the TBRE hamiltonian.

\vspace{0.2in} FIG. 3 ~~  The predicted $I$g.s. probabilities of
four very different systems. Solid squares are obtained by diagonalizing
the TBRE hamiltonian  and open squares are the predicted
$I$g.s. probabilities of this Letter. a) $j=\frac{9}{2}$ with 4
fermions; b) $j=\frac{9}{2}$ with 5 fermions; c) a 7 fermions in
a two-$j$ shell ($j_1=7/2, j_2=5/2$); d) an $sd$-boson system with
10 bosons.

\newpage

{TABLE I. The angular momenta which give the lowest eigenvalues
when $G_J=-1$  and all other parameters are  0 for 4 fermions in
single-$j$ shells.  }

\begin{footnotesize}
\begin{tabular}{ccccccccccccccccc} \hline  \hline
$2j$ &  $G_0$ &  $G_2$ &  $G_4$ &  $G_6$ &  $G_8$ &  $G_{10}$ &  $G_{12}$
&  $G_{14}$ &  $G_{16}$ &  $G_{18}$ &  $G_{20}$ &  $G_{22}$ &  $G_{24}$
&  $G_{26}$ &  $G_{28}$ &  $G_{30}$   \\  \hline
7  & 0 &4 &2 &8 &   &   & & & &  & & & & & & \\
9  & 0 &4 &0 &0 &12 &   & & & &  & & & & & & \\
11 & 0 &4 &0 &4 &8  &16 & & & &  & & & & & & \\
13 & 0 &4 &0 &2 &2  &12 &20 & & &  & & & & & & \\
15 & 0 &4 &0 &2 &0  &0  &16 &24 & &  & & & & & & \\
17 & 0 &4 &6 &0 &4  &2  &0  &20 &28 &  & & & & & & \\
19 & 0 &4 &8 &0 &2  &8  &2  &16 &24 &32 & & & & & & \\
21 & 0 &4 &8 &0 &2  &0  &0  &0  &20 &28 &36 & & & & & \\
23 & 0 &4 &8 &0 &2  &0  &10 &2  &0  &24 &32 &40 & & & & \\
25 & 0 &4 &8 &0 &2  &4  &8  &10 &6  &0  &28 &36 &44 & & & \\
27 & 0 &4 &8 &0 &2  &4  &2  &0  &0  &4  &20 &32 &40 &48 & & \\
29 & 0 &4 &8 &0 &0  &2  &6  &8  &12 &8  &0  &24 &36 &44 &52 & \\
31 & 0 &4 &8 &0 &0  &2  &0  &8  &14 &16 &6  &0  &32 &40 &48 & 56 \\
\\   \hline  \hline
\end{tabular}
\end{footnotesize}


\newpage


\begin{thebibliography}{50}
\bibitem{Johnson1} C. W. Johnson, G. F. Bertsch, D. J. Dean, Phys. Rev. Lett.
{\bf 80}, 2749(1998).

\bibitem{Johnson2}C. W. Johnson, G. F. Bertsch, D. J. Dean, and I. Talmi,
Phys. Rev. C{\bf 61}, 014311(1999).

\bibitem{Johnson3} C. W. Johnson, Rev. Mex. Fix. {\bf
45 ~ S2}, 25(1999).

\bibitem{Bijker}
R. Bijker, A. Frank, Phys.  Rev. Lett. {\bf 84}, 420(2000);
Phys. Rev. C{\bf 62}, 14303(2000).


\bibitem{Casten}
D. Kusnezov, N. V. Zamfir, and R. F. Casten,
 Phys.  Rev. Lett. {\bf 85}, 1396(2000).


\bibitem{Horoi} M. Horoi, B. A. Brown, V. Zelevinsky,
Phys. Rev. Lett. {\bf 87}, 062501(2001).


\bibitem{Zhao0}Y. M. Zhao, A. Arima, and N. Yoshinaga, to be published.

\bibitem{Zuker} V. Velazquez, and A. P. Zuker, nucl-th/0106020.


\bibitem{Bijker1} R. Bijker, A. Frank, and S. Pittel,
Phys.  Rev. C{\bf 60}, 021302(1999).


\bibitem{zelevinsky}D. Mulhall, A. Volya, and V. Zelevinsky,
Phys. Rev. Lett.85, 4016(2000); Nucl. Phys. {\bf A682}, 229c(2001);
V. Zelevinsky et al., Yad. Fiz. {\bf 64}, 579(2001).

\bibitem{Zhaox} Y.M. Zhao,  and A. Arima,  Phys. Rev. {\bf C64}, (R)041301(2001).


\bibitem{Kus} D. Kusnezov, Phys. Rev. Lett.  {\bf 85}, 3773(2000);
ibid. {\bf 87}, 029202 (2001);
R. Bijker, and A. Frank, Phys. Rev. Lett. {\bf 87}, 029201(2001).


\bibitem{Bijkerx} R. Bijker, and A. Frank, nucl-th/0105027; nucl-th/0108054;
nucl-th/0111041; Phys. Rev. {\bf C64}, 061303(R)(2001).

\bibitem{Johnsonx} Lev Kaplan, Thomas Papenbrock, and Calvin W. Johnson,
Phys. Rev. {\bf C63}, 014307(2001);   Lev Kaplan, and Thomas Papenbrock,
Phys. Rev. Lett. {\bf 84}, 4553(2000).


\bibitem{note1} The lowest
eigenvalues  should be  equivalent to the largest
eigenvalues  in our discussion. But,
these largest eigenvalues are usually (exactly or nearly) zero
for many $I$ matrices, especially for many-$j$ shells and large single-$j$
shells. To have the ``rule" as simple as possible, we
use only the lowest eigenvalues with one of $G_J=-1$ and others
switched off.


\bibitem{note2}
The  term $ (s^{\dagger} d^{\dagger}) (s d)$ gives degenerate
eigenvalues  for many $I$ states. Therefore,
we may also use 6 (instead of 7)  as the number of
independent two-body interactions.
The difference due to this minor modification is very small, though.



\end{thebibliography}
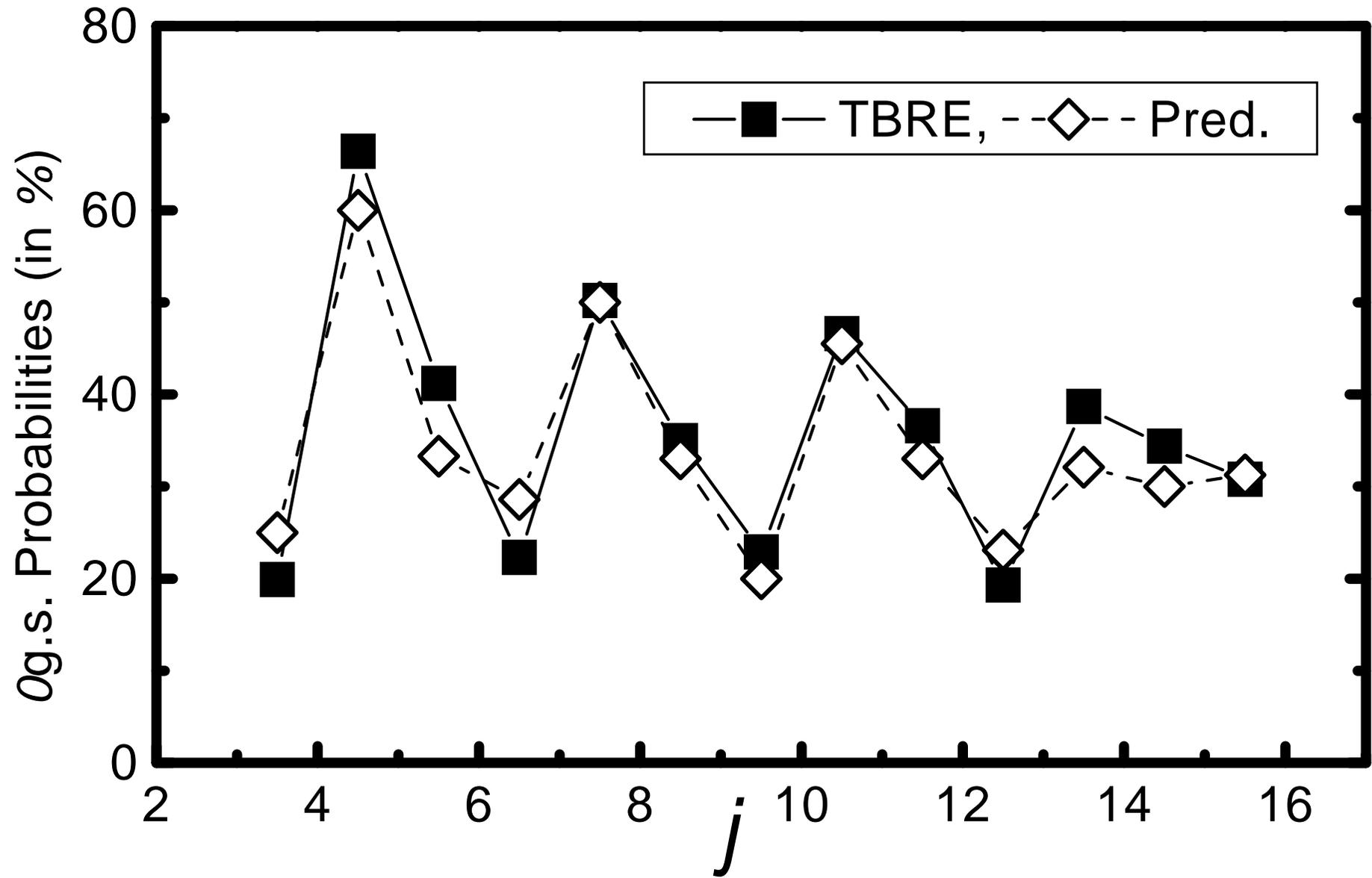
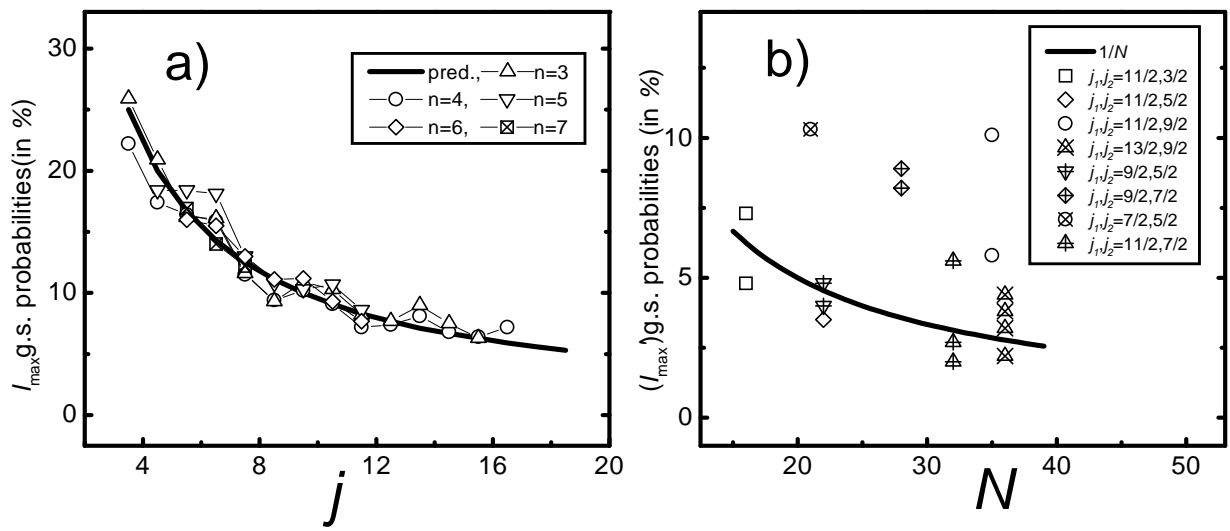
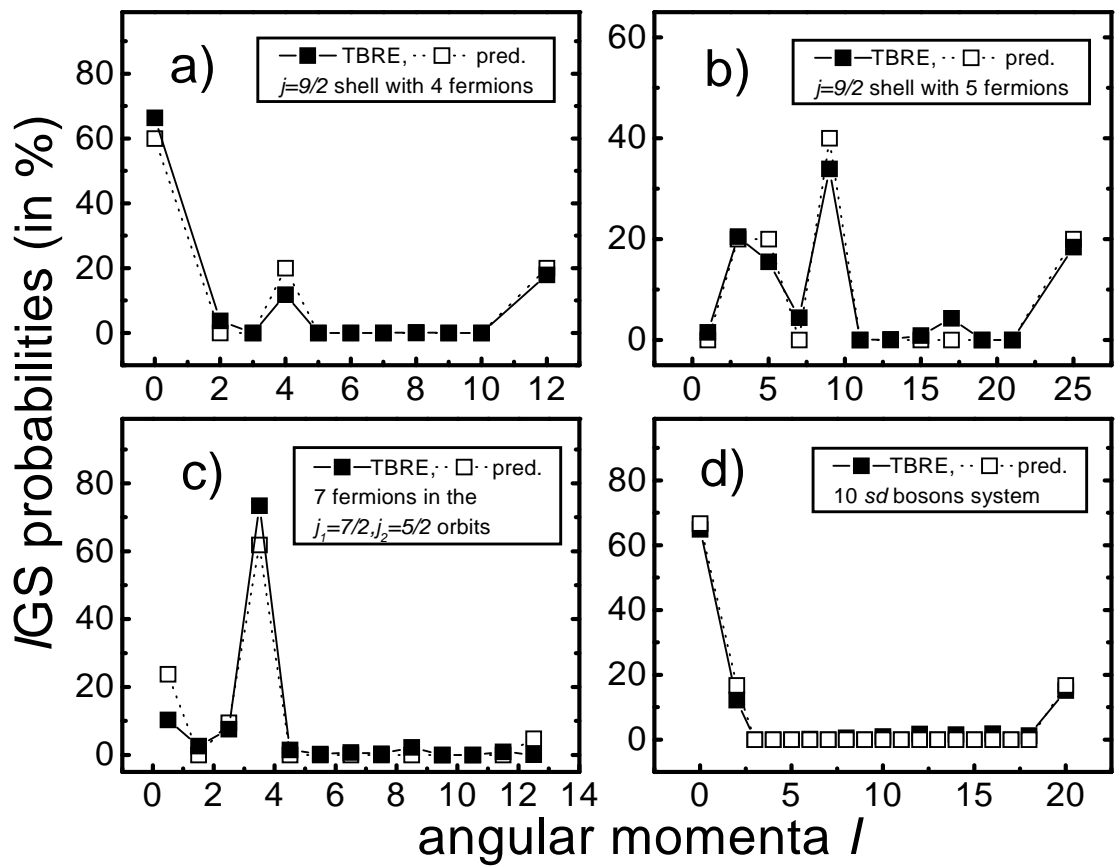
\end{document}